\newcommand{\be}{\begin{equation}} \newcommand{\ee}{\end{equation}}
\def\bea{\begin{eqnarray}}
\def\eea{\end{eqnarray}}

\documentstyle{article}
\begin{document}
\begin{flushright}    UFIFT-HEP-98-23 
%hep-ph/9809401 
\end{flushright}
\vskip 2cm
\centerline{\bf Neutrinos: A Glimpse Beyond the Standard Model}
\vskip .5cm
\centerline{P. Ramond}
\vskip .2cm

\centerline{Institute for Fundamental Theory, 
Department of Physics} 
\centerline{University of Florida, Gainesville, Fl 32611}
\vskip .5cm
\centerline{(Neutrino-98, Takayama, Japan, June 1998)}

%\begin{abstract}
%\end{abstract}
%\maketitle
\vskip .5cm
\centerline{\it Dedicated to the Memory of Dick Slansky}
\vskip 1cm
\section{A Short History of Neutrinos}
Neutrinos are awesome: of all elementary particles, only neutrinos
(not even quarks!) have their own conferences, this year Neutrino-98, 
on a par with Susy, Strings, Lattices, and the like.

It is sobering  to remind ourselves that all weak interaction
 experiments start out wrong, even when performed by the greatest 
experimentalists of their times. 
In 1911-1912, using a magnetic spectrometer and photographic plates, 
O. Von Bayer, O. Hahn, and L. Meitner~\cite{RAD,VOL} were the first 
to measure the spectrum of electrons in $\beta$ radioactivity. Their 
conclusion: like $\alpha$ radioactivity, the spectrum of the decay 
product is discrete!

In 1914, Chadwick~\cite{CHAD}, 
performed similar measurements in Geiger's laboratory in Berlin 
and came out with a different conclusion, that the spectrum of 
$\beta$ electrons is continuous. The Great War interrupted the discourse, 
and the next step in the story were measurements by
C. D. Ellis~\cite{ELLIS} who showed that the discrete lines found
earlier were due to internal conversion. Finally in 1927, C.D. 
Ellis and W. A. Wooster~\cite{ELWO} found that the mean energy 
liberated in $\beta$ 
decay accounted for only $1/3$ of the allowed 
energy. By that time even Lise Meitner agreed 
that the electron spectrum was continuous, 
setting the stage for W. Pauli's famous letter. 

In a December 1930 letter that starts with typical panache, $``${\it  
Dear Radioactive Ladies and Gentlemen...}", W. Pauli proposes 
a $``${\it desperate}"way out: there is a companion particle 
to the $\beta$ electron. Undetected, it must be electrically 
neutral, and in order to balance the $N-Li^6$ statistics, it 
carries spin $1/2$. He calls it the {\it neutron}. It is clear 
from the letter that Pauli saw no reason why this new particle 
could not be massive. 

In 1933, E. Fermi in his formulation of the theory of $\beta$ decay
gave it its final name, the little neutron or {\it neutrino}, as it is
clearly much lighter than Chadwick's neutron which had been discovered
since Pauli's letter. 

The next step in our story is in 1945, when B. Pontecorvo~\cite{PONTA}
puts forward the idea that neutrinos can be detected. It is based on
the following
observation: an electron neutrino can hit a ${^{37}Cl}$ atom and
transform it into ${^{37}Ar}$. While the Chlorine atoms are plentiful, as in cleaning fluid $C_2Cl_4$, Argon is an inert gas
that does not interact much; furthermore it is  radioactive and
sticks around just long enough to be detectable through its decay: 
its abundance  can be monitored
by patient and careful experimentalists.  Pontecorvo did not publish
the report, perhaps because of its secret classification, or perhaps because
he showed it to Fermi  who  thought the idea
ingenious but not immediately  achievable. 

In 1953,  Cowan and  Reines~\cite{COREA} proposed a different
technique to detect neutrinos,  by using a liquid scintilator. 

In 1954,  Davis~\cite{DAVISA} uses Pontecorvo's original
proposal, by setting up outside a nuclear reactor, and then using
radio-chemical techniques to detect the Argon atoms. 

In 1956, Cowan and Reines~\cite{COREB} announced they had detected $\overline \nu_e$'s 
 through the reaction
$\overline \nu_e+p\rightarrow e^+ +n$. Cowan passed away before 1995,
the year Fred Reines was awarded the Nobel Prize for their discovery. 
There emerge two  lessons in neutrino physics: not only is patience
required but also longevity: it took $26$ years from birth to
detection and then another $39$ for the Nobel Committee to recognize 
the achievement!

In 1956, motivated by rumors that Davis had found evidence for
antineutrinos coming from a pile, Pontecorvo~\cite{PONTB} reasoned, in
analogy to  Gell-Mann and Pais, who  had just shown how a $K$-meson could
oscillate into its antiparticle, that it could be due to a similar effect: an
electron neutrino produced in the Savannah reactor could oscillate
into its own antiparticle and be detected by Davis. The rumor went
away, but the idea of neutrino oscillations was born; it has remained
with us ever since, and proven the most potent tool in hunting for
neutrino masses. 

Having detected the neutrino, there remained to determine its spin and
mass. Its helicity was measured in 1958 by M. Goldhaber~\cite{MGOLD},
but convincing evidence  for its mass has, up to this meeting, eluded
experimentalists. 

In 1957, Lee and Yang propose that weak interactions violate parity, and the neutrino is again at the center of the action. Unlike the charged elementary particles which have both left- and right-handed components, neutrinos are purely left-handed (antine
utrinos are right-handed), which means that lepton-number is chiral. 

In 1962, a second neutrino, the muon neutrino  is detected~\cite{2NEUT}, (long anticipated by theorists Inou\"e and Sakata in 1943~\cite{INSA}). This time things went a bit faster as it took only 19 years from theory (1943) to discovery (1962) and 26 year
s to Nobel recognition (1988). 

That same year, Maki, Nakagawa and Sakata ~\cite{MANASA} introduce two crucial ideas; one is that these two neutrinos can mix, and the second is that this mixing can cause one type of neutrino to oscillate into the other (called today flavor oscillation).
 This is possible only if the two neutrino flavors have different masses.

In 1963, the Astrophysics group at Caltech, Bahcall, Fowler,
Iben and Sears~\cite{BWIS} puts forward the most accurate of neutrino
fluxes from the Sun. Their calculations included the all important
Boron decay spectrum, which produces neutrinos with the right energy
range for the Chlorine experiment. 

In 1964, using Bahcall's result~\cite{BAH} of an enhanced capture rate of ${^8B}$ neutrinos through an excited state of ${^{37}Ar}$, Davis~\cite{DAVISB}
proposes to search for ${^8B}$ solar neutrinos using a $100,000$ gallon
tank of cleaning fluid deep underground.  Soon after, R. Davis starts 
his epochal experiment at the Homestake mine, marking the
beginning of the solar neutrino watch which continues to this day. In
1968, Davis et al reported~\cite{DAVISC} a deficit in the solar neutrino flux, a result that has withstood scrutiny to this day, and stands as a truly
remarkable experimental {\it tour de force}. Shortly after, Gribov and
Pontecorvo~\cite{GRIPO} interpreted the deficit as evidence for neutrino oscillations.

\section{Standard Model Neutrinos}
The standard model of electro-weak and strong interactions contains three left-handed neutrinos.  The three neutrinos are represented by two-components Weyl spinors, $\nu^{}_{i}$, $i=e,\mu,\tau$, each describing a left-handed fermion (right-handed antifer
mion). As the upper components of weak isodoublets $L^{}_i$, they have $I^{}_{3W}=1/2$, and a unit of the global $i$th lepton number. 

These standard model neutrinos are strictly massless. The only Lorentz scalar made out of these neutrinos is the Majorana mass, of the form
$\nu^{t}_{i}\nu^{}_{j}$; it has the quantum numbers of a weak isotriplet, with third component  $I^{}_{3W}=1$, as well as two units of total lepton number. Thus to generate a Majorana mass term at tree-level, one  needs a Higgs isotriplet with two units 
 of lepton number. Since the standard model Higgs is a 
 weak isodoublet Higgs,  there are no tree-level neutrino masses.

What about quantum corrections? Their effects are  not limited to
renormalizable couplings, and it is easy to make a weak isotriplet out
of two isodoublets, yielding the $SU(2)\times U(1)$ invariant
$L^t_i\vec\tau L^{}_j\cdot H^t_{}\vec\tau H$, where $H$ is the Higgs
doublet. As  this term is not invariant under lepton number, it is not be generated in perturbation theory. Thus the
important conclusion: {\it The standard model neutrinos are kept
massless by global chiral lepton number symmetry}. The detection of non-zero neutrino masses is a tangible indication of physics beyond the standard model.
\section{Neutrino Mass Models}
The present experimental limits on neutrino masses are quite impressive, $m_{\nu_e}< 
10~ {\rm eV}$, $m_{\nu_\mu}< 170~ {\rm keV}$, $m_{\nu_\tau}< 18~ {\rm MeV}$~\cite{PDG}.  Any model that generates neutrino masses must contain a natural mechanism that explains their small value, relative to that of their charged counterparts. 

To generate neutrino masses without new fermions, we must break lepton number. This  requires adding to the standard model 
 Higgs fields which carry lepton number, as one can arrange  to 
break lepton number explicitly or spontaneously through their interactions.  

To impart Higgs with lepton number, they must be coupled to standard
model leptons. From invariance requirements, we see that there can be
only three such fields with two units of lepton number: An isotriplet 
Higgs, $\vec{\bf T}$, and two isosinglets, one 
positively charged,  $S_{}^+$, the other  doubly charged, $S_{}^{--}$, 
with renormalizable couplings
\be \vec{\bf T}\cdot L_{(i}\vec\tau L_{j)}\ ;\qquad S_{}^+L_{[i}\vec\tau
L_{j]}\ ;\qquad S_{}^{--}\overline e^{}_{(i}\overline e^{}_{j)}\ .\ee
The curvy brackets denote flavor-symmetrization, the square ones  
flavor antisymmetrization. 

With these fields we can construct three types of  cubic interactions
that break lepton number: $H\vec\tau H\cdot \vec{\bf T}$, $S_{}^+ S_{}^+S_{}^{--}$, and
$\vec{\bf T}\cdot \vec{\bf T} S_{}^{--}$, which  introduce through their couplings 
an unknown scale at which lepton number is violated. There are no
quartic interactions that violate lepton number. 

The Higgs isotriplet has a neutral compent; it can be arranged to get
a vacuum value, breaking lepton number spontaneously. This leads
to a Nambu-Goldstone boson, called the Majoron. Since it is part of an
isotriplet, it couples to the Z boson, whose measured width rules out isotriplet breaking
of lepton number. One needs electroweak singlet scalars with  lepton number 
to devise  Majoron models that are not in manifest conflict with experiment.

Perhaps the simplest way to give neutrinos masses is to introduce for
each one an electroweak singlet Dirac partner, $\overline
N^{}_i$. These appear naturally in the Grand Unified group
$SO(10)$. Neutrino Dirac masses are generated by the couplings $L^{}_i\overline 
N^{}_j H$ after electroweak breaking. Unfortunately, these Yukawa
couplings yield masses which are too big: they are along the electroweak breaking
parameter, of the same order of magnitude as the masses of the charged
elementary particles $m\sim\Delta
I_w=1/2$  . The situation is remedied by introducing
Majorana mass terms $\overline N^{}_i\overline N^{}_j$ for the
right-handed neutrinos. The masses of these new degrees of freedom is
arbitrary, since it has no electroweak quantum numbers, $M\sim\Delta
I_w=0$. If it is much larger than the electroweak scale, the
neutrino masses are suppressed relative to that of their charged
counterparts by the ratio of the electroweak scale to that new scale: 
the mass matrix  (in $3\times 3$ block form) is
\be
\hskip 1in \pmatrix{0& m\cr m&M}\ ,
\ee
leading to one small and one large eigenvalue 
\be
m_\nu~\sim~ m\cdot {m\over M}~\sim~ \left(\Delta I_w={1\over 2}\right)\cdot 
\left({ \Delta I_w={1\over 2}
\over \Delta I_w=0 }\right)\ .\ee 
This seesaw mechanism~\cite{SEESAW} provides a natural
explanation for the smallness of the neutrino masses as long as lepton
number is broken at a large scale $M$. With $M$ around the energy at
which the gauge couplings unify, this yields neutrino masses at or
below the eV region. 

The flavor mixing comes from two different parts, the diagonalization
of the charged lepton Yukawa couplings, and that of the neutrino
masses. From the charged lepton Yukawas, we obtain ${\cal U}_e^{}$, 
the unitary matrix that rotates the
lepton doublets $L^{}_i$. From the neutrino Majorana matrix, we obtain
$\cal U_\nu$, the matrix that diagonalizes the Majorana mass matrix. 
The $6\times 6$ seesaw Majorana matrix can be written in $3\times 3$
block form
\be
{\cal M}={\cal V}_\nu^t ~{\cal D} {\cal V}^{}_\nu\sim\pmatrix {{\cal
U}_\nu&\epsilon {\cal U}^{}_{N\nu}\cr
\epsilon{\cal U}^{t}_{N\nu}&{\cal U}^{}_{N}\cr}\ ,\ee
where $\epsilon$ is the tiny rastio of the electroweak to lepton
number violating scales, and ${\cal D}={\rm diag}(\epsilon^2{\cal D}_\nu, {\cal D}_N)$,
 is a diagonal matrix. ${\cal D}_\nu$ contains the
three neutrino masses, and $\epsilon^2$ is the seesaw suppression. The
weak charged current is then given by
\be
j^+_\mu=e^\dagger_i\sigma_\mu {\cal U}^{ij}_{MNS}\nu_j\ ,\ee
where
\be
{\cal U}^{}_{MNS}={\cal U}^{}_e{\cal U}^\dagger_\nu\ ,\ee
is the matrix first introduced in ref [14], the analog of the CKM
matrix in the quark sector. 

In the seesaw-augmented standard model, this mixing matrix is totally 
arbitrary. It contains, as does the CKM matrix, three rotation angles,
and one CP-violating phase, and also two additional CP-violating phases
which cannot be absorbed in a redefinition of the neutrino
fields, because of their Majorana masses (these extra phases can be measured only in $\Delta {\cal L}=2$ processes). All are additional
parameters of the seesaw-augmented standard model, to be determined by
experiment.

Their prediction, as for for the quark hierarchies and mixings,
necessitates further theoretical assumptions. Below we present such a
framework, which predicts maximal mixing between $\nu_\mu$ and
$\nu_\tau$~\cite{HRR} and a thrice Cabibbo suppression of $\nu_e$ into
$\nu_{\mu,\tau}$. 

\section{A Neutrino Mixing Model}
This model~\cite{ILR} follows from the Cabibbo suppresions of the Yukawa couplings of
the standard model. Using the well-known Cabibbo
suppressions in the quark sector, we identify family symmetries on the
quarks that reproduce the patterns. We generalize this symmetry to the
leptons, using grand-unified groups in a very simple way, and then use
the lepton assignments to produce Cabibbo suppressions in the lepton
sectors. Using special properties of the seesaw mechanism, we find 
 a unique lepton mixing matrix, with the properties
already described.  

We assume that the Cabibbo supression comes about because of extra
family symmetries in the standard model. A standard model invariant
operator, such as ${\bf Q}_i{\bf \overline d}_jH_d$, if not 
invariant under the additional symmetry, cannot be present
at tree-level. Assuming the existence of an electroweak singlet field $\theta$, which serves as the order paramameter for this new
symmetry, the interaction
\be
{\bf Q}_i{\bf \overline d}_jH_d\left({\theta\over \Lambda}\right)^{n_{ij}}\ee
can appear in the potential as long as the family charges balance under the
new symmetry. When $\theta$ acquires a $vev$, this leads to a
suppression of the Yukawa couplings of the order of $\lambda^{n_{ij}}$
for each matrix element, where
$\lambda=\theta/\Lambda$ is assumed to be like the Cabibbo angle, and
$\Lambda$ is the natural cut-off of the theory.  This is a natural
mechanism in the context of an effective low energy theory with cut-off
$\Lambda$. As a consequence of the charge balance equation
\be X_{if}^{[d]}+n^{}_{ij}X^{}_\theta=0\ ,\ee
the exponents of the suppression is related to the charge of the
standard model invariant operator. That charge is the sum of the
charges of the fields that make up the invariant. Let us now apply
this mechanism to the invariants in the seesaw mechanism. 

We start with the charged lepton Yukawa couplings of the form
$L_i\overline N_j H_u$, with charges $X_{L_i}+X_{Nj}+X_{H}$, which
gives the Cabibbo suppression of the $ij$ matrix element. It follows
that we can write the orders of magnitude of these couplings in the
form
\be
\pmatrix{\lambda^{l_1}&0&0\cr
0&\lambda^{l_2}&0\cr
0&0&\lambda^{l_3}\cr}{\cal Y}\pmatrix{\lambda^{p_1}&0&0\cr
0&\lambda^{p_2}&0\cr
0&0&\lambda^{p_3}\cr}\ ,\ee
where ${\cal Y}$ is a Yukawa matrix with no Cabibbo
suppressions, $l_i=X_{L_i}/X_\theta$,
$p_i=X_{N_i}/X_\theta$. The first matrix will form the first half of
the MNS matrix in the charged lepton current. 

Similarly, the mass matrix for the right-handed
neutrinos, $\overline N_i\overline N_j$ will be written in the form
\be
\pmatrix{\lambda^{p_1}&0&0\cr
0&\lambda^{p_2}&0\cr
0&0&\lambda^{p_3}\cr}{\cal M}\pmatrix{\lambda^{p_1}&0&0\cr
0&\lambda^{p_2}&0\cr
0&0&\lambda^{p_3}\cr}\ .\ee
The diagonalization of the  seesaw matrix is of the form 
\be 
L_iH_u\overline N_j \left({1\over{{\overline N}~\overline
N}}\right)_{jk}\overline N_kH_uL_l\ ,\ee
from which the Cabibbo suppression matrix from the $\overline N_i$
fields {\it cancels}, leaving us with
\be
 \pmatrix{\lambda^{l_1}&0&0\cr
0&\lambda^{l_2}&0\cr
0&0&\lambda^{l_3}\cr}{\cal M}'\pmatrix{\lambda^{l_1}&0&0\cr
0&\lambda^{l_2}&0\cr
0&0&\lambda^{l_3}\cr}\ ,\ee
where ${\cal M}'$ is a matrix with elements of order one. The Cabibbo structure of the seesaw neutrino matrix is determined
solely by the charges of the lepton doublets! As a result, the Cabibbo
structure of the MNS
mixing matrix is also due entirely to the charges of the three lepton
doublets. This general conclusion depends on the existence of at least
one Abelian family symmetry, which we argue is implied by the observed
structure in the quark sector.

The Wolfenstein parametrization of the CKM matrix~\cite{WOLF}, 
\be
\hskip 1in
\pmatrix{1&\lambda & \lambda ^3\cr
             \lambda &1&\lambda ^2\cr 
             \lambda ^3&\lambda ^2&1}\ ,\ee
and the Cabibbo
structure of the quark mass ratios
\be {m_{u}\over m_t}\sim \lambda ^8\;\;\;{m_c\over m_t}\sim 
\lambda ^4\;\;\; ;
\;\;\; {m_d\over m_b}\sim \lambda ^4\;\;\;{m_s\over m_b}\sim
\lambda^2\ ,\ee
are reproduced by a simple charge assignment on the three quark families, namely
\be
X_{{\bf Q},{\bf \overline u},{\bf \overline d}} ={\cal B}(2,-1,-1)+
\eta_{{\bf Q},{\bf \overline u},{\bf \overline d}}(1,0,-1)\ ,\ee
where ${\cal B}$ is baryon number, $\eta_{{\bf \overline d}}=0 $, and 
$\eta_{{\bf Q}}=\eta_{{\bf \overline u}}=2$. 
Two noteworthy  features emerge: the charges of the down quarks 
associated with the second and third families are the same, and the
$\eta$ values for both ${\bf Q}$ and ${\bf \overline u}$ are the same.

Theoretical prejudices based on grand unified quantum numbers  determine for us the family charges of the leptons from those of the quarks. In grand unified extensions of the standard model,  
baryon number generalizes in $SO(10)$ to ${\cal B}-{\cal L}$, where ${\cal L}$ is total
lepton number, and  the standard model families split under
$SU(5)$ as  ${\bf\overline 5}={\bf
\overline d}+L$, and ${\bf 10}={\bf Q}+{\bf \overline u}+\overline
e$. Thus a natural assignment is
to assign $\eta=0$ to the lepton doublet $L_i$, and $\eta=2$ to the
electron singlet $\overline e_i$. In this way, the charges of the
lepton doublets are simply $X_{L_i}=-1(2,-1,-1)$. As we have just
argued, these charges determine the Cabibbo structure of the MNS
mixing matrix to be
\be
\hskip .2in
{\cal U}^{}_{MNS}\sim\pmatrix{{\cal O}(1)&{\cal O}(\lambda^3)&{\cal O}(\lambda^3)\cr
{\cal O}(\lambda^3)&{\cal O}(1)&{\cal O}(1)\cr {\cal O}(\lambda^3)&{\cal O}(1)&{\cal O}(1)\cr}\ .\ee
We therefore expect no Cabibbo suppression in the mixing between
$\nu_\mu$ and $\nu_\tau$. This mixing scheme is consistent with the preliminary
results of SuperKamiokande announced at the 1977 ITP workshop~\cite{SUPERK}, and
also consistent with the small angle MSW~\cite{MSW} solution to the solar
neutrino deficit. 

The determination of the mass values is more complicated, as it not
only depends on the relative interfamily charge assignments but also
on the overall intrafamily charges. Here we simply quote the results
from a particular model~\cite{ILR}.  
The masses of the right-handed neutrinos are found to be of the
following orders of magnitude
\be
m_{\overline N_e}\sim \Lambda\lambda^{13}\ ;\qquad m_{\overline N_\mu}\sim
m_{\overline N_\tau}\sim \Lambda\lambda^7\ ,\ee
where $\Lambda$ is the cut-off. The seesaw mass matrix for the three light  neutrinos comes out to be 
\be
\hskip .5in
 m^{}_0\pmatrix{a\lambda^6&b\lambda^3&c\lambda^3\cr
b\lambda^3&d&e\cr
c\lambda^3&e&f\cr}\ ,\ee
where we have added for future reference the prefactors $a,b,c,d,e,f$, all of order one, and 
\be m_0^{}={v_u^2\over
{\Lambda\lambda^3}}\ ,\ee
where $v_u$ is the $vev$ of the Higgs doublet. This matrix has one light eigenvalue
\be
m_{\nu_e}\sim m_0^{}\lambda^6_{}\ .\ee
Without a detailed analysis of the prefactors, the masses of the other two neutrinos come out  to be both of 
 order $m_0$. 
However, the mass difference inferred by the superKamiokande
result~\cite{SUPERK} (up to this conference) can be reproduced, but only
if the prefactors are carefully taken into account.  The two heavier mass
eigenstates and their mixing angle are written in terms of 
\be
x={df-e^2\over (d+f)^2}\ ,\qquad y={d-f\over d+f}\ ,\ee
as
\be {m_{\nu_2}\over m_{\nu_3}}={1-\sqrt{1-4x}\over 1+\sqrt{1-4x}}\
,\qquad \sin^22\theta_{\mu\tau}=1-{y^2\over 1-4x}\ .\ee
If $4x\sim 1$, the two heaviest neutrinos are nearly degenerate. If
$4x\ll 1$, a condition easy to achieve if $d$ and $f$ have the same
sign, we can obtain an adequate split between the two mass
eigenstates. For illustrative purposes, when $0.03<x<0.15$, we find
\be
4.4\times 10^{-6}\le \Delta m^2_{\nu_e-\nu_\mu}\le 10^{-5}~{rm eV}^2\
  ,\ee
which yields the correct non-adiabatic MSW effect, and
\be
5\times 10^{-4}\le  \Delta m^2_{\nu_\mu-\nu_\tau}\le 5\times
10^{-3}~{\rm eV}^2\ ,\ee
for the atmospheric neutrino effect. These were calculated with a
cut-off, $10^{16}~{\rm GeV}<\Lambda<4\times 10^{17}~{\rm GeV}$, and a mixing
angle, $0.9<\sin^22\theta_{\mu-\tau}<1$. It is satisfying that these
values are compatible not only with the data but also with the gauge
unification scale, and the basic ideas of Grand Unification. With 
poetic justice, we note that Grand Unification with its prediction 
of proton decay motivated the building of large underground water 
\v Cerenkov counters. The serendipitous detection of neutrinos from
SN1987A  by the IMB, Kamiokande,  and 
other collaborations, established these detectors as major tools for the discovery 
of neutrino properties. 
\section{Outlook}
The present field of neutrino physics is being driven by many
experimental findings that challenge theoretical
expectations. All can be explained in terms of neutrino oscillations,
implying neutrino masses and mixing angles, but one should be cautious
as evidence for neutrino oscillations has often been reported, 
only to either be withdrawn or else contradicted by other
experiments. 

The reported anomalies associated with solar neutrinos~\cite{SOLAR},
neutrinos produced in cosmic ray cascades~\cite{SUPERK}, and also in low energy
reactions~\cite{LSND},  cannot all be correct without introducing a new
type of neutrino which does not couple to the $Z$ boson, a {\it
sterile}  neutrino.

Small neutrino masses are
naturally generated by the seesaw mechanism, which works because of
the weak interactions of the neutrinos.  A similar mass suppression 
for sterile neutrinos involves new hitherto unknown interactions,
resulting in substantial additions to the standard model, for which
there is no independent evidence. Also, the case for a
heavier cosmological neutrino in aiding structure formation may not be as pressing, in
view of the measurements of a small cosmological constant.

To conclude, experimental neutrino physics is in a most exciting stage, as it 
provides  in the near future the best opportunities for finding evidence of 
physics beyond the standard model.

\end{document}